# Results for grundy number of the complement of bipartite graphs

Ali Mansouri ,Mohamed Salim Bouhlel
Department of Electronic Technologies of Information and Telecommunications Sfax,
Tunisia.
mehermansouri@yahoo.fr
medsalim.bouhlel@enis.rnu.tn

**ABSTRACT**

Our work becomes integrated into the general problem of the stability of the network ad hoc. Some, works attacked(affected) this problem. Among these works, we find the modelling of the network ad hoc in the form of a graph. Thus the problem of stability of the network ad hoc which corresponds to a problem of allocation of frequency amounts to a problem of allocation of colors in the vertex of graph. we present use a parameter of coloring " the number of Grundyö

A Grundy k-coloring of a graph G, is a vertex k-coloring of G such that for each two colors i and j with i < j, every vertex of G colored by j has a neighbor with color i. The Grundy chromatic number (G), is the largest integer k for which there exists a Grundy k-coloring for G. In this note we Łrst give an interpretation of (G) in terms of the total graph of G, when G is the complement of a bipartite graph. Then we prove that determining the Grundy number of the complement of bipartite graphs is an NP-Complete problem.

## 1 INTRODUCTION

In this paper we consider undirected graphs without loops and multiple edges. By a k-coloring of a graph G we mean a proper vertex coloring of G with colors1,2,...,k.
A Grundy k-coloring of G is a k-coloring of G such that for each two colors i and j with i < j, every vertex of G colored by j has a neighbor with color i. The Grundy chromatic number (or simply Grundy number) (G), is the largest integer k for which there exists a Grundy k-coloring for G.
The Grundy number of graphs was perhaps introduced for the Łrst time by Christen and Selkow [2]. In [3] another interpretation of Grundy number has been obtained.
Also in [7] the authors studied Grundy number of hypercubes and determined the exact values. Further results on Grundy numbers have been found in [11].
From computational point of view, in[6]a linear algorithm for determining (T) has Been given, where T is a tree. Also in[9]a polynomial time algorithm for computing Grundy number of partial k-trees has been found. In the unpublished manuscript [4] the NP-Completeness of determining Grundy number of general graphs has been proved.
.

## 2 GRUNDY NUMBER

The Grundy number of a graph G is the maximum number k of colors used to color the vertices of G such that the coloring is proper and every vertex x colored with color i, 1 Ö i Ö k, is adjacent to (i - 1) vertices colored with each color j, 1 Öj Öi ó 1.

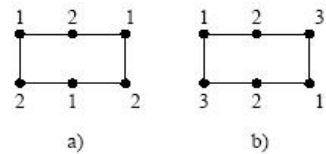

**Figure 1:** coloration de grundy.

**Instance**: A graph G and an integer k.
**Question**: Is (G)× k?
In this paper our aim is to study Grundy number of the complement of bipartite graphs and give a description of it in terms of total graphs. Finally we will prove the NP-Completeness of Grundy number for this restricted class of graphs. Suppose G is the complement of a bipartite graph with a bipartition (X,Y). By an extended clique of G we mean a

subgraph of G, which can be defined inductively as follows:

An ordinary clique of G is also an extended clique of G. Then, a subgraph H is an extended clique if there exist two non-adjacent vertices $u \in X$ and $v \in Y$ in H such that $H \setminus \{u,v\}$ is an extended clique of G. An extended clique in G, in fact introduces an independent subset I of vertices in the bipartite graph $G^c$, and a matching in $G^c \setminus I$ with m vertices. By the size of this extended clique we mean the number $|I|+m$.

In this note the concept of total graph has been used. The total graph T(G) of a graph G, is the graph whose vertices correspond to the vertices and edges of G, and where two vertices are joined if and only if the corresponding vertices or edges of G are adjacent. The total graph T(G) of G has the following property. Suppose a property P for a graph G is defined on the set $V(G) \cup E(G)$. Then this property can be converted into a property P′ only on the vertices of T(G). Total graphs were defined for the first time in [1]. We need also the concept of an edge dominating set. In a graph G a subset D of edges in G is called an edge dominating set if each edge in $E(G)\setminus D$ has a common end point with an edge in D. By the size of D we simply mean the number of edges in D. In our NP-Completeness result we have transformed the following result of Yannakakis and Gavril [10] concerning the minimum size of an edge dominating set in any graph G.

**Theorem A**. Determining the minimum size of an edge dominating set in a bipartite graph with maximum degree at most 3, is an NP-Complete problem.

## 3 RESULTAT

An extended clique of G with size t in fact introduces a Grundy coloring of G with at least t colors. The following theorem shows that the converse is also true.

**Theorem 1**. Let G be the complement of a bipartite graph. Then there exists an extended clique in G with size $\Gamma(G)$.[12]

**Proof.** Suppose that $\Gamma(G) = t$. We prove by induction on t that an extended clique in G with size t can be obtained by a Grundy coloring with t colors in G.

We may suppose that the class of vertices colored by 1 in a Grundy t-coloring of G consists of two vertices. Otherwise, if any color class consists of one vertex then clearly $\Gamma(G) = t$, where $\omega(G)$ is the size of a maximum clique in G, and in this case G should be a complete graph for which the theorem holds. Therefore suppose that a color class say the class i contains two vertices and let i be the minimum color having this property. Then by changing the classes 1 and i with each other, we will have a Grundy t-coloring of G with the desired property. Now if we consider two vertices of color 1 and delete them, then the resulting graph H has Grundy number $t-1$. By induction on t, we may suppose that the Grundy number $t-1$ for H is obtained by an extended clique in H having size $t-1$. Now by adding two vertices colored by 1 to that extended clique in H, we obtain an extended clique with size t in G, as required.

The theorem in fact shows that if G is the complement of a bipartite graph then $\Gamma(G)$ equals to the maximum size of an extended clique in G.

In the following theorem, by $\alpha(G)$ we mean the maximum number of independent vertices in a graph G.

**Theorem 2**. Suppose G is a bipartite graph. Then

$$\Gamma(G^c) = \alpha(T(G)). \qquad (1)$$

**Proof**. The Grundy number of $G^c$ is the maximum size of an extended clique in $G^c$ and this later size can be taken as the maximum size of a set K of vertices and edges in G where both vertices and edges are independent and no edge in K is adjacent to a vertex in K. But such a set K in G introduces an independent subset in T(G), and vice versa any independent set in T(G) provides an extended clique in $G^c$. Therefore $\Gamma(G^c)$ is the same as the maximum number of independent vertices in T(G), namely $\alpha(T(G))$.

**Theorem 3**. Let G be the complement of a bipartite graph. Then

$$\chi(G) \leq 3\Gamma(G)/2 \qquad (2)$$

**Proof**. The Grundy number $\Gamma(G)$ is the size of some extended clique in G. The later size is of the form $m+|I|$, where m is the size of a matching M in $G^c$ and I an independent subset of vertices in $G^c$ which are not incident to any edge in M. Let $|I| = a+b$, $a \geq b$, where a is the number of vertices in I belonging to one unique partite in $G^c$. Let us denote the independence number of $G^c$ by $\alpha$. It is clear that $a \leq \alpha/2$ and $b+m \leq \alpha$. Consequently

$$\chi(G) = a+b+m \leq \alpha +a \leq 3\alpha/2 = 3\Gamma(G)/2 \qquad (3)$$

The above theorem implies the existence of a good polynomial time approximation algorithm for Grundy number with a performance ratio 3/2.

Before we state the next theorem, we mention the following fact from edge dominating sets in bipartite graphs. Let D be an edge dominating set in a bipartite graph, then there is a matching M which is also an edge dominating set and $|M| \leq |D|$.

**Theorem 4.** The following decision problem is NP-Complete:

**Instance:** A bipartite graph G with $\Delta(G) \leq 3$ and an integer k.

**Question:** Is $\Gamma(G^c) \geq k$?

**Proof**. A proper vertex coloring of $G^c$ can be easily checked whether it is a Grundy coloring with at least k number of colors. This shows the membership in NP. We transform EDGE DOMINATING SET (EDS) problem to our problem. Suppose $(G, k)$ is an instance for EDS. We show that $(G^c, n-k)$ is an appropriate instance for GRUNDY NUMBER, where n is the order of G. Let D be an edge dominating set in G with a tmost k edges. Then there exists a matching M which is also an edge dominating set and $|M| \leq |D|$. Suppose that $M = \{v_1u_1, v_2u_2, \cdots, v_mu_m\}$. Let $V_0$ be those vertices in G which are not incident with any edge in M. Now it turns out that $V_0$ in conjunction with the pairs $\{v_i, u_i\}, 1 \leq i \leq m$, forms an extended clique in $G^c$. Its size is $n-m$. Now if $m \leq k$, then $n-m \geq n-k$. Similarly in the reverse procedure, an extended clique in $G^c$ of size m introduces an edge dominating set in G of size $n-m$. This completes the proof.

## ACKNOWLEDGEMENT

I think Mr. Manouchehr Zaker for his collaboration.